\documentclass[12pt]{iopart}
\usepackage[dvipdfmx]{graphicx}
\usepackage{xcolor}

\begin{document}

\title[A heralded single-photon source implemented with nonlinear photonic crystals]{A heralded single-photon source implemented with second-order nonlinear photonic crystals}

\author{Hiroo Azuma}

\address{Global Research Center for Quantum Information Science, National Institute of Informatics, 2-1-2 Hitotsubashi, Chiyoda-ku, Tokyo 101-8430, Japan}
\ead{zuma@nii.ac.jp}
\vspace{10pt}
\begin{indented}
\item[]March 2025
\end{indented}

\begin{abstract}
We study implementing a heralded single-photon source with second-order nonlinear photonic crystals.
Injecting pump and signal light beams into a one-dimensional photonic crystal composed of a material with a large second-order nonlinear optical susceptibility $\chi^{(2)}$,
we can transform the coherent incident signal light into squeezed light.
Preparing two squeezed light beams by this method
and input them into two ports of a 50-50 beam splitter independently,
we can transform them into a two-mode squeezed state.
Because those two modes of photons share entanglement,
a single photon of the one mode is emitted with a high probability
on the condition of the single-photon detection of the other mode as a heralding signal.
We evaluate the efficiency and the second-order intensity correlation function $g^{(2)}(0)$ for this heralded single-photon source.
Moreover, we examine changes in the efficiency and $g^{(2)}(0)$ when we assume that the single-photon detector for the heralding signal is imperfect.
For a specific concrete example, we consider a case where the nonlinear medium is lithium niobate $\mbox{LiNbO}_{3}$
and the frequency of the incident signal light is on the order of $10^{14}$ Hz.
\end{abstract}

%
%
%
%
%

\section{\label{section-Introduction}Introduction}
An on-demand single-photon source is a key component for realization of the quantum information processing.
We need to prepare single-photon sources to perform
Knill, Laflamme, and Milburn's controlled two-qubit gate \cite{Shen2018},
the quantum cryptography protocol BB84 (Bennett-Brassard 1984) \cite{Rusca2018},
the quantum teleportation \cite{Hu2020},
the quantum secure direct communication \cite{Deng2004,Ying2024,Liu2025,Ying2025},
the measurement-device-independent quantum key distribution \cite{Lo2012},
the blind quantum computation \cite{Fitzsimons2017,Sheng2018},
and the entanglement concentrarion \cite{Sheng2012a,Sheng2012b}.
So far, many devices for the realization of the on-demand single-photon gun have been proposed.
Those proposals include methods, for example,
using cavity quantum electrodynamics
\cite{Mucke2013,Bergmann2019,Meher2022},
quantum dots
\cite{Senellart2017,Zhu2022},
and
nitrogen vacancy centers in diamond
\cite{Dreau2018,Bathen2021}.
Despite all those efforts, the ideal and practical implementation of the single-photon source has not been established yet.

As an alternative to implementing the on-demand single-photon gun,
we can construct a heralded single-photon source.
The most typical method for the alternative is utilization of spontaneous parametric down-conversion (SPDC)
\cite{Ngah2015,Kaneda2015,Zhang2021,Dhara2022}.
The SPDC emits
a pair of photons in two different directions.
Thus, if we detect one photon as the heralding signal,
the other one flies out at the time of the detector's measurement.
We can regard this phenomenon as a practical process for the deterministic single-photon source.
However, the SPDC has a disadvantage that the generation probability of the entangled pairs of photons is considerably low.
It was reported that the conversion efficiency was $4\times 10^{-6}$ per pump photon
at most when using a periodically poled lithium niobate (PPLN) waveguide for the SPDC
\cite{Bock2016}.

How to demonstrate experiments of the heralded single-photon source is a currently ongoing study.
References~\cite{Davis2022,Stasi2023} proposed improvement of the heralded single-photon source using a high-efficiency photon-number-resolving detector.
The authors of Ref.~\cite{Tang2021} investigated the utilization of the photon-blockade effect for reducing multiphoton events caused by the SPDC
with a strong pump power and overcoming the purity-yield trade-off of the heralded single-photon source.

In this paper, we propose a method to implement the heralded single-photon source with second-order nonlinear photonic crystals.
This method consists of the following three stages.
First, we inject pump and signal light beams into a one-dimensional photonic crystal composed of a material with a large second-order nonlinear optical susceptibility $\chi^{(2)}$.
Then, the incident coherent signal light is transformed into squeezed light.
Second, we prepare two squeezed light beams by the first stage and inject them into two ports of a 50-50 beam splitter independently.
Then, the beam splitter transforms those incident beams into a two-mode squeezed state.
Third, with the detection of a single photon of one mode as a heralding signal,
a heralded single photon of the other mode is emitted from the beam splitter with a high probability.
For a concrete example, we assume
that the photonic crystal is composed of air and lithium niobate $\mbox{LiNbO}_{3}$.
Our numerical estimations show that the probability
that we can obtain the heralded single photon is larger than $0.2$ per pump photon.
Thus, we can conclude that our proposal is superior to the heralded single-photon source realized by the SPDC.

To judge whether or not the physical properties of our photon source are close
to those of an ideal on-demand single-photon source,
we use the second-order intensity correlation function $g^{(2)}(0)$
\cite{Walls2008}.
In general, $0\leq g^{(2)}(0)$ holds,
and we obtain $g^{(2)}(0)=0$ when an on-demand identical photon gun is realized.
We show that $g^{(2)}(0)$ for the proposed source can be nearly equal to zero
when we give some specific physical parameters.

Our proposal relies on the following two facts:
\begin{enumerate}
\item[(a)]
The photonic crystal forces the group velocity of the incident light to decrease.
We can derive this characteristic by using the classical electrodynamics,
i.e., Maxwell's equations.
However, we can use this classical characteristic of the photonic crystal to amplify the quantum nature of a wave function of the photons.
The slow group velocity makes the traveling time of the photons in the photonic crystal larger.
Thus, if the photonic crystal is composed of a nonlinear material with the susceptibility $\chi^{(2)}$,
the time duration of nonlinear interaction between the photons and the material increases
and we can obtain a strong nonlinear effect.
\item[(b)]
We inject two squeezed light beams into two ports of the 50-50 beam splitter independently.
Then, the beam splitter transforms them into the two-mode squeezed state.
This two-mode squeezed state promises that our single-photon source is high quality,
that is, $0<g^{(2)}(0)\ll 1$.
In Ref.~\cite{Azuma2022},
only one squeezed state was injected into the 50-50 beam splitter
to generate another two-mode squeezed state
but it did not attain
small $g^{(2)}(0)$.
\end{enumerate}

The realization of slow light in the photonic crystal was studied both theoretically and experimentally
\cite{Baba2008,Torrijos-Moran2021,Lu2022}.
Fabrication of the photonic crystal from $\mbox{LiNbO}_{3}$ attracts a lot of attention from many researchers in the field of photonic integrated circuits
\cite{Li2020,Jiang2020,Larocque2024}.
Lithium niobate $\mbox{LiNbO}_{3}$ is a typical material that has a large second-order nonlinear optical susceptibility $\chi^{(2)}$.

This paper is organized as follows.
In Sec.~\ref{section-review-photonic-crystal},
we review how to generate the squeezed light with the photonic crystal.
In Sec.~\ref{section-heralded-single-photon-source},
we propose an optical circuit that works as the heralded single-photon source.
In Sec.~\ref{section-functions-optical-circuit},
we explain the functions of the optical circuit.
In Sec.~\ref{section-perfect-g-2-function},
we compute the efficiency and $g^{(2)}(0)$
of the proposed heralded single-photon source using a perfect single-photon detector.
In Sec.~\ref{section-imperfect-single-photon-detector},
we evaluate the efficiency and $g^{(2)}(0)$ of the heralded single-photon source with an imperfect single-photon detector.
In Sec.~\ref{section-photonic-crystal-group-velocity-squeezing-parameter},
we calculate the group velocity of the photon passing the photonic crystal and the squeezing parameter
induced by the nonlinearity of $\mbox{LiNbO}_{3}$.
In Sec.~\ref{section-conclusions-discussions},
we provide conclusions and discussions.
In \ref{section-photonic-crystal-length-calibration},
we compute the ratio of probabilities that the photons stay in the layers of the photonic crystal.

\section{\label{section-review-photonic-crystal}A brief review of how to generate squeezed light
with the nonlinear photonic crystal}
\begin{figure}
\begin{center}
\includegraphics[width=0.65\linewidth]{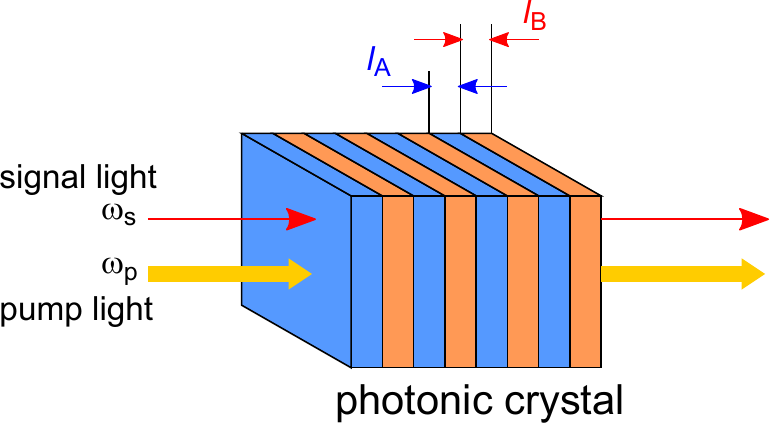}
\end{center}
\caption{A schematic of a one-dimensional nonlinear photonic crystal,
and incident pump and signal light beams.
The angular frequencies of the signal and pump light beams are given by $\omega_{\mbox{\scriptsize s}}$ and $\omega_{\mbox{\scriptsize p}}$, respectively,
with $\omega_{\mbox{\scriptsize p}}=2\omega_{\mbox{\scriptsize s}}$.}
\label{figure01}
\end{figure}

In this section, we review how to squeeze coherent light with a nonlinear photonic crystal
\cite{Azuma2022,Sakoda2005}.
As shown in Fig.~\ref{figure01}, we construct the nonlinear photonic crystal by periodically depositing the layers of two different materials A and B
that have different linear refractive indices in a one-dimensional stack
with widths of $l_{\mbox{\scriptsize A}}$ and $l_{\mbox{\scriptsize B}}$, respectively.
For a concrete example, we assume that materials A and B are air and $\mbox{LiNbO}_{3}$,
respectively.
As mentioned in Sec.~\ref{section-Introduction},
$\mbox{LiNbO}_{3}$ has a large second-order nonlinear optical susceptibility $\chi^{(2)}$.
For the sake of simplicity,
we suppose that $\chi^{(2)}$ is real and positive.
As indicated in Fig.~\ref{figure01}, we inject the pump and signal light beams
with angular frequencies $\omega_{\mbox{\scriptsize p}}$ and $\omega_{\mbox{\scriptsize s}}(=\omega_{\mbox{\scriptsize p}}/2)$, respectively,
into the photonic crystal.

We assume that the electric field of the pump light is sufficiently strong
so that we consider the pump light to be classical without quantum fluctuations.
Contrastingly, because the amplitude of the signal light is very weak,
we need to consider it as a quantum one.
We assume that the pump and signal light beams propagate in $x$-direction
and they are polarized in $y$-direction.
Thus, the wave vectors of the pump and signal light beams are given by
$\mbox{\boldmath $k$}_{\mbox{\scriptsize p}}=(k_{\mbox{\scriptsize p}},0,0)$
and
$\mbox{\boldmath $k$}_{\mbox{\scriptsize s}}=(k_{\mbox{\scriptsize s}},0,0)$
with $k_{\mbox{\scriptsize p}}=2k_{\mbox{\scriptsize s}}$.
We can express the electric fields of the pump and signal light beams as
$\mbox{\boldmath $E$}_{\mbox{\scriptsize p}}=(0,E_{\mbox{\scriptsize p}},0)$,
$\hat{\mbox{\boldmath $E$}}_{\mbox{\scriptsize s}}=(0,\hat{E}_{\mbox{\scriptsize s}},0)$,
\begin{equation}
E_{\mbox{\scriptsize p}}(x,t)
=
iA
\{
\exp[i(k_{\mbox{\scriptsize p}}x-\omega_{\mbox{\scriptsize p}}t+\theta)
-
\exp[-i(k_{\mbox{\scriptsize p}}x-\omega_{\mbox{\scriptsize p}}t+\theta)
\},
\label{pump-electric-field-0}
\end{equation}
\begin{equation}
\hat{E}_{\mbox{\scriptsize s}}(x,t)
=
i
\sqrt{\frac{\hbar\omega_{\mbox{\scriptsize s}}}{2\varepsilon_{0}V}}
\{
\exp[i(k_{\mbox{\scriptsize s}}x-\omega_{\mbox{\scriptsize s}}t)]
\hat{a}_{\mbox{\scriptsize s}}
-
\exp[-i(k_{\mbox{\scriptsize s}}x-\omega_{\mbox{\scriptsize s}}t)]
\hat{a}_{\mbox{\scriptsize s}}^{\dagger}
\},
\label{signal-light-electric-field-0}
\end{equation}
where $\theta$, $A$, and $V$ represent the phase of the pump light,
the amplitude of the pump light,
and
the volume for quantization of the signal light, respectively.
In Eq.~(\ref{signal-light-electric-field-0}),
$\varepsilon_{0}$ denotes vacuum permittivity,
$\hbar$ is equal to $h/(2\pi)$,
$h$ represents the Planck constant,
and $\hat{a}_{\mbox{\scriptsize s}}$ and $\hat{a}_{\mbox{\scriptsize s}}^{\dagger}$
represent annihilation and creation operators of the signal light, respectively.
Signal light that has passed the photonic crystal of width $l$ is described by the form:
\begin{equation}
\hat{E}_{\mbox{\scriptsize s}}(l,t)
=
i
\sqrt{\frac{\hbar\omega_{\mbox{\scriptsize s}}}{2\varepsilon_{0}V}}
\{
\exp[i(k_{\mbox{\scriptsize s}}l-\omega_{\mbox{\scriptsize s}}t)]
\hat{b}_{\mbox{\scriptsize s}}(l)
-
\exp[-i(k_{\mbox{\scriptsize s}}l-\omega_{\mbox{\scriptsize s}}t)]
\hat{b}_{\mbox{\scriptsize s}}^{\dagger}(l)
\},
\end{equation}
\begin{equation}
\hat{b}_{\mbox{\scriptsize s}}(l)
=
\cosh(\beta l)\hat{a}_{\mbox{\scriptsize s}}
+
e^{i\theta}\sinh(\beta l)\hat{a}_{\mbox{\scriptsize s}}^{\dagger},
\end{equation}
\begin{equation}
\beta
=
\frac{\omega_{\mbox{\scriptsize s}}A\chi^{(2)}}{\varepsilon_{0}v_{\mbox{\scriptsize g}}},
\label{definition-beta-parameter-0}
\end{equation}
where $v_{\mbox{\scriptsize g}}$ denotes the group velocity of the signal light in the photonic crystal.
An operator $\hat{b}_{\mbox{\scriptsize s}}(l)$ is obtained
by the Bogoliubov transform of $\hat{a}_{\mbox{\scriptsize s}}$.
Introducing a squeezing parameter and a squeezing operator:
\begin{equation}
\zeta=\beta l e^{i\theta},
\label{definition-squeezing-parameter-0}
\end{equation}
\begin{equation}
\hat{S}(\zeta)
=
\exp
\left(
-
\frac{\zeta}{2}\hat{a}_{\mbox{\scriptsize s}}^{\dagger 2}
+
\frac{\zeta^{*}}{2}\hat{a}_{\mbox{\scriptsize s}}^{2}
\right).
\end{equation}
we obtain the following relation:
\begin{equation}
\hat{S}(\zeta)\hat{a}_{\mbox{\scriptsize s}}\hat{S}^{\dagger}(\zeta)
=
\hat{b}_{\mbox{\scriptsize s}}(l).
\end{equation}
If we place a coherent state $|\alpha\rangle$ as the signal light into the photonic crystal,
it emits a squeezed state $|\zeta,\alpha\rangle$, where
\begin{equation}
|\zeta,\alpha\rangle
=
\hat{S}(\zeta)|\alpha\rangle.
\end{equation}

\section{\label{section-heralded-single-photon-source}An optical circuit for the heralded single-photon source}
\begin{figure}
\begin{center}
\includegraphics[width=0.65\linewidth]{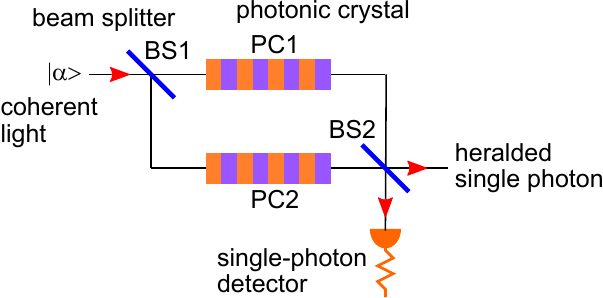}
\end{center}
\caption{An optical circuit for generating a two-mode squeezed state with the photonic crystals
and beam splitters and emitting a single photon by detection of a heralded single photon.
Here, we indicate that the two-mode squeezed state was used for constructing the heralded single-photon source in past literature.
The generation of the two-mode squeezed state in atomic mechanical oscillators was studied in \cite{Leong2023}.
Implementation of the heralded single-photon source using the two-mode squeezed state with an imperfect photon detector
was studied theoretically in \cite{Sekatski2012} and experimentally in \cite{Sempere-Llagostera2022}.
}
\label{figure02}
\end{figure}

Figure~\ref{figure02} shows an optical circuit for the generation of the heralded single photon.
First, we inject a beam of coherent light $|\alpha\rangle$ into the beam splitter BS1.
Then, two beams of coherent light are emitted from two ports of BS1 separately.
Second, we inject those two beams of coherent light into the photonic crystals PC1 and PC2
as the signal light beams separately.
Here, we omit to draw the pump light beams in Fig.~\ref{figure02}.
Then, the photonic crystals PC1 and PC2 apply the Bogoliubov transform to the two incident beams of coherent light and emit two beams of squeezed light.
Third, we inject those two beams of the squeezed light
into two ports of the beam splitter BS2 independently.
Finally, beams of a two-mode squeezed state are emitted from two ports of the beam splitter BS2.
Those two modes are entangled with each other.
Thus, if we detect a single photon of the one mode as a heralding signal
with the single-photon detector,
we obtain an emission of a heralded single photon of the other mode with a high probability.
Hence, we can conclude that the optical circuit shown in Fig.~\ref{figure02}
realizes the heralded single-photon source.

\section{\label{section-functions-optical-circuit}Functions of the optical circuit}
We follow the progress of the photonic state step by step
when we inject coherent light $|\alpha\rangle$
into the optical circuit shown in Fig.~\ref{figure02}.
First, the beam splitter BS1 splits the coherent light $|\alpha\rangle$ into two beams.
Here, we describe the annihilation and creation operators of photons of the two input ports of the beam splitter as
$\hat{a}$, $\hat{a}^{\dagger}$, $\hat{b}$, and $\hat{b}^{\dagger}$, respectively.
A 50-50 beam splitter applies the following unitary operator to the incident photons:
\begin{equation}
\hat{B}(\delta)
=
\exp
\left[
\frac{\pi}{4}(e^{i\delta}\hat{a}^{\dagger}\hat{b}-e^{-i\delta}\hat{a}\hat{b}^{\dagger})
\right].
\end{equation}
We assume that the unitary transformation of BS1 is given by $\hat{B}(0)$.
The displacement operator $\hat{D}(\alpha)$ that generates the coherent light $|\alpha\rangle$ satisfy the following relations:
\begin{equation}
|\alpha\rangle_{a}
=
\hat{D}_{a}|0\rangle_{a},
\end{equation}
\begin{equation}
\hat{D}_{a}(\alpha)
=
\exp(\alpha\hat{a}^{\dagger}-\alpha^{*}\hat{a}),
\quad
\hat{D}_{b}(\alpha)
=
\exp(\alpha\hat{b}^{\dagger}-\alpha^{*}\hat{b}),
\end{equation}
\begin{equation}
\hat{B}(0)\hat{D}_{a}(\alpha)\hat{B}^{\dagger}(0)
=
\hat{D}_{a}(\alpha/\sqrt{2})
\hat{D}_{b}(-\alpha/\sqrt{2}),
\end{equation}
\begin{equation}
\hat{B}(0)|\alpha\rangle_{a}
=
|\alpha/\sqrt{2}\rangle_{a}|-\alpha/\sqrt{2}\rangle_{b}.
\end{equation}
Thus, the injection of the coherent light $|\alpha\rangle_{a}$ into BS1
causes emissions of two coherent light beams
$|\alpha/\sqrt{2}\rangle_{a}$
and
$|-\alpha/\sqrt{2}\rangle_{b}$ from ports $a$ and $b$, respectively.

Second, we inject those coherent light beams into the nonlinear photonic crystals
PC1 and PC2 separately.
Letting $\theta=0$ for Eq.~(\ref{pump-electric-field-0}),
we can write the Bogoliubov transform given rise to by the nonlinear photonic crystal as the squeezing operator $\hat{S}(r)$ with $r=\beta l$.
Accordingly, the following state is emitted from PC1 and PC2:
\begin{equation}
|r,\alpha/\sqrt{2}\rangle_{a}|r,-\alpha/\sqrt{2}\rangle_{b}
=
\left[
\hat{S}_{a}(r)|\alpha/\sqrt{2}\rangle_{a}
\right]
\left[
\hat{S}_{b}(r)|-\alpha/\sqrt{2}\rangle_{b}
\right].
\end{equation}

Third, we inject the above two squeezed light beams into two ports of the beam splitter BS2 independently.
We assume that the unitary operator BS2 is given by $\hat{B}(\pi/2)$.
Then, the squeezing operators of the two modes are transformed as follows:
\begin{eqnarray}
&&
[\hat{B}(\pi/2)\hat{S}_{a}(r)\hat{B}^{\dagger}(\pi/2)]
[\hat{B}(\pi/2)\hat{S}_{b}(r)\hat{B}^{\dagger}(\pi/2)] \nonumber \\
&=&
\hat{S}_{ab}(ir/2)\hat{S}_{a}(r/2)\hat{S}_{b}(-r/2)
\hat{S}_{a}(-r/2)\hat{S}_{b}(r/2)\hat{S}_{ab}(ir/2) \nonumber \\
&=&
\hat{S}_{ab}(ir),
\end{eqnarray}
where
\begin{equation}
\hat{S}_{ab}(\zeta)
=
\exp(-\zeta\hat{a}^{\dagger}\hat{b}^{\dagger}+\zeta^{*}\hat{a}\hat{b}).
\end{equation}
Moreover, the displacement operators of the two modes are transformed as follows:
\begin{eqnarray}
\hat{B}(\pi/2)\hat{D}_{a}(\alpha/\sqrt{2})\hat{B}^{\dagger}(\pi/2)
&=&
\hat{D}_{a}(\alpha/2)\hat{D}_{b}(i\alpha/2), \nonumber \\
\hat{B}(\pi/2)\hat{D}_{b}(-\alpha/\sqrt{2})\hat{B}^{\dagger}(\pi/2)
&=&
\hat{D}_{a}(-i\alpha/2)\hat{D}_{b}(-\alpha/2).
\end{eqnarray}
Thus, the final state emitted from the optical circuit in Fig.~\ref{figure02} is given by
\begin{equation}
\hat{S}_{ab}(ir)
\hat{D}_{a}
\left(
\frac{\alpha}{2}(1-i)
\right)
\hat{D}_{b}
\left(
\frac{\alpha}{2}(i-1)
\right)
|0\rangle_{a}|0\rangle_{b}.
\label{two-mode-squeezed-state-with-alpha}
\end{equation}

\section{\label{section-perfect-g-2-function}The efficiency and the second-order intensity correlation function
$g^{(2)}(0)$ of the heralded single-photon source
for the case where the single-photon detector is perfect}
In this section, we numerically compute the efficiency and the second-order intensity correlation function $g^{(2)}(0)$
of the heralded single-photon source
for the case where the single-photon detector is perfect.
In other words, we assume that the single-photon detector can perfectly distinguish
an event of a single photon
from the zero-photon event and the events that include two or more photons.
The final state emitted from the optical circuit in Fig.~\ref{figure02} is given by
\begin{equation}
\hat{S}_{ab}(ir)
\hat{D}_{a}
(
\beta
)
\hat{D}_{b}
(
-\beta
)
|0\rangle_{a}|0\rangle_{b},
\end{equation}
where
$\beta=(\alpha/2)(1-i)$.
The probability that we detect $n_{a}$ and $n_{b}$ photons in the ports $a$ and $b$ of BS2,
respectively, is written in the form
\begin{equation}
P(n_{a},n_{b};r,\alpha)
=
|
\mu(n_{a},n_{b};r,\alpha)
|^{2}
\label{probability-na-nb-perfect-0}
\end{equation}
where
\begin{eqnarray}
&&
\mu(n_{a},n_{b};r,\alpha) \nonumber \\
&=&
{}_{a}\langle n_{a}|_{b}\langle n_{b}|
\hat{S}_{ab}(ir)
\hat{D}_{a}(\beta)
\hat{D}_{b}(-\beta)
|0\rangle_{a}|0\rangle_{b} \nonumber \\
&=&
\sum_{m_{a}=0}^{\infty}
\sum_{m_{b}=0}^{\infty}
[{}_{ab}\langle n_{a}, n_{b}|
\hat{S}_{ab}(ir)
|m_{a},m_{b}\rangle_{ab}]
[{}_{a}\langle m_{a}|\hat{D}_{a}(\beta)|0\rangle_{a}] \nonumber \\
&& \times
[{}_{b}\langle m_{b}|\hat{D}_{b}(-\beta)|0\rangle_{b}].
\end{eqnarray}
The elements of the matrices,
$
{}_{ab}\langle n_{a}, n_{b}|
\hat{S}_{ab}(ir)
|m_{a},m_{b}\rangle_{ab}
$,
${}_{a}\langle m_{a}|\hat{D}_{a}(\beta)|0\rangle_{a}$,
\\
and
${}_{b}\langle m_{b}|\hat{D}_{b}(-\beta)|0\rangle_{b}$
are given by
\begin{eqnarray}
&&
{}_{ab}\langle n_{a},n_{b}|
\hat{S}_{ab}(\zeta)
|m_{a},m_{b}\rangle_{ab} \nonumber \\
&=&
\sum_{n=0}^{\mbox{\scriptsize min}[n_{a},n_{b}]}
\sum_{m=0}^{\mbox{\scriptsize min}[m_{a},m_{b}]}
\delta_{n_{a}-n,m_{a}-m}
\delta_{n_{b}-n,m_{b}-m}
\frac{1}{n!m!}
(-1)^{n}
e^{i(n-m)\varphi}
\tanh^{n+m}r \nonumber \\
&&
\times
\exp[-(m_{a}+m_{b}-2m+1)\ln(\cosh r)]
\frac{\sqrt{n_{a}!n_{b}!m_{a}!m_{b}!}}{(n_{a}-n)!(n_{b}-n)!},
\label{matrix-elements-S-ab}
\end{eqnarray}
\begin{equation}
{}_{a}\langle m_{a}|\hat{D}_{a}(\beta)|0\rangle_{a}
=
\exp
\left(
-\frac{1}{2}|\beta|^{2}
\right)
\frac{\beta^{m_{a}}}{\sqrt{m_{a}!}},
\end{equation}
\begin{equation}
{}_{b}\langle m_{b}|\hat{D}_{b}(-\beta)|0\rangle_{b}
=
\exp
\left(
-\frac{1}{2}|\beta|^{2}
\right)
\frac{(-\beta)^{m_{b}}}{\sqrt{m_{b}!}},
\end{equation}
respectively,
where
$\zeta=r e^{i\varphi}$.

Using $P(n_{a},n_{b};r,\alpha)$ given by
Eq.~(\ref{probability-na-nb-perfect-0}),
we compute $g^{(2)}(0)$ numerically.
Writing $n$ as the number of the emitted photons with the single-photon heralding signal,
we can calculate the average of $n^{l}$ by
\begin{equation}
\langle n^{l}\rangle
=
\sum_{n=0}^{\infty}
n^{l}
P(n,1;r,\alpha).
\label{average-n-l-0}
\end{equation}
Using Eq.~(\ref{average-n-l-0}), we can compute $g^{(2)}(0)$ as
\begin{equation}
g^{(2)}(0)
=
\frac{\langle n^{2}\rangle-\langle n\rangle}{\langle n\rangle^{2}}.
\label{g2-0-definition}
\end{equation}

\begin{figure}
\begin{center}
\includegraphics[width=0.65\linewidth]{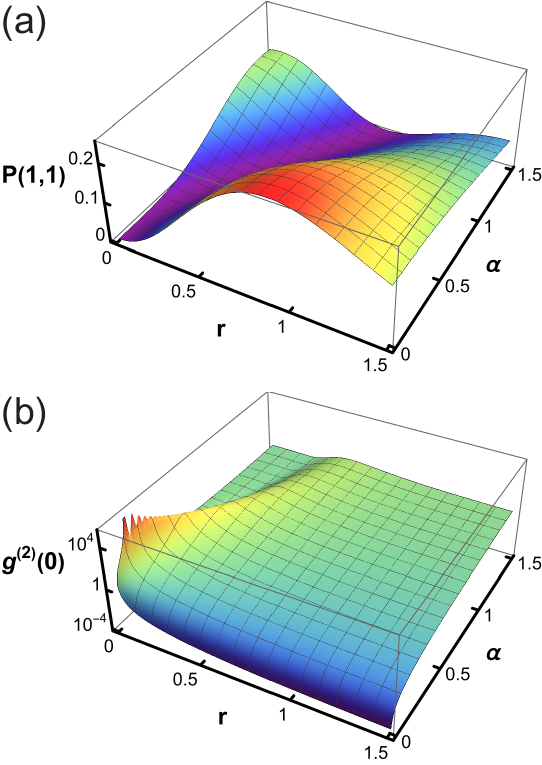}
\end{center}
\caption{(a)
A 3D plot of $P(1,1;r,\alpha)$ given by Eq.~(\ref{probability-na-nb-perfect-0})
as a function of $r$ and $\alpha$.
When $\alpha=0$, $P(1,1;r,\alpha)$ attains the maximum value $1/4$
for $r=\log(1+\sqrt{2})\approx 0.881$.
(b)
A 3D plot of $g^{(2)}(0)$ given by Eqs.~(\ref{average-n-l-0}) and (\ref{g2-0-definition})
as a function of $r$ and $\alpha$.
The vertical axis is displayed by the logarithmic scale.
Looking at this graph, we note that $g^{(2)}(0)$ takes a large value when $r\ll 1$ and $\alpha<0.5$.
However, $g^{(2)}(0)=1$ must hold for $r=0$ because the state of photons in the port $a$ is given by a coherent state with $r=0$.
This fact implies that $g^{(2)}(0)$ has a singularity at $r=0$ if we regard $g^{(2)}(0)$ as a function of $r$.
We can find the origin of this singularity in Eq.~(\ref{matrix-elements-S-ab}).
Equation~(\ref{matrix-elements-S-ab}) includes terms $\tanh^{n+m}r$.
Because of $\tanh r=0$ for $r=0$, the terms $\tanh^{n+m}r$ cause the singularity of $0^{0}$.
}
\label{figure03}
\end{figure}

\begin{figure}
\begin{center}
\includegraphics[width=0.65\linewidth]{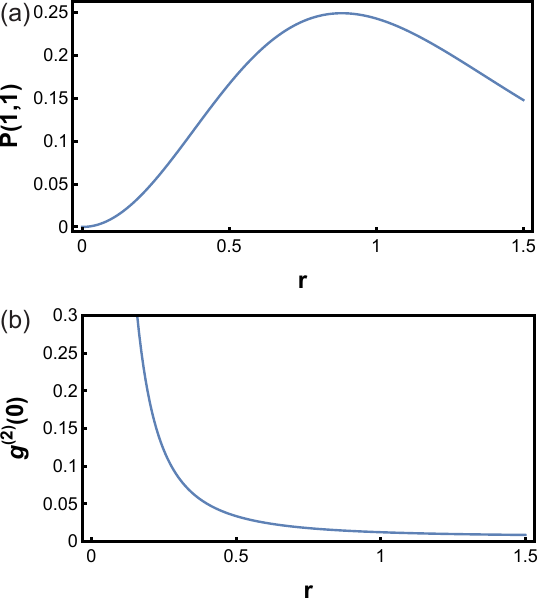}
\end{center}
\caption{
(a) A plot of $P(1,1;r,\alpha)$ given by Eq.~(\ref{probability-na-nb-perfect-0})
as a function of $r$ with $\alpha=0.06$.
The value of $P(1,1;r,0.06)$ attains the maximum value $0.249$ for $r=0.883$.
(b) A plot of $g^{(2)}(0)$ given by Eqs.~(\ref{average-n-l-0}) and (\ref{g2-0-definition})
as a function of $r$ with $\alpha=0.06$.
When $r=1.5$, we obtain $g^{(2)}(0)=8.78\times 10^{-3}$.}
\label{figure04}
\end{figure}

Figure~\ref{figure03}(a) shows $P(1,1;r,\alpha)$ as a function of $r$ and $\alpha$
for
$0\leq r\leq 1.5$ and $0\leq\alpha\leq 1.5$.
We can regard $P(1,1;r,\alpha)$ as the efficiency of the heralded single-photon source.
Looking at Fig.~\ref{figure03}(a),
we note that $P(1,1;r,\alpha)$ achieves more than $0.2$ for $r \approx 0.881$ and $0\leq\alpha\leq 0.4$.
Thus, we can consider that the efficiency of the heralded single-photon source is larger than $0.2$ per pump photon.
Figure~\ref{figure03}(b) shows a 3D plot of $g^{(2)}(0)$ as a function of $r$ and $\alpha$.
Looking at Fig.~\ref{figure03}(b), we note that $0<g^{(2)}(0)<0.1$ for $0.5\leq r\leq 1.5$ and $0\leq\alpha\leq 0.1$.
As $\alpha$ gets larger or $r$ gets smaller, $g^{(2)}(0)$ increases.

Figure~\ref{figure04}(a) shows a plot of $P(1,1;r,\alpha)$ as a function of $r$ with $\alpha=0.06$.
We obtain $P(1,1;r=0.883,\alpha=0.06)=0.249$
and $P(1,1;r=1.5,\alpha=0.06)=0.148$.
Figure~\ref{figure04}(b) shows a plot of $g^{(2)}(0)$ as a function of $r$ with $\alpha=0.06$.
We obtain $g^{(2)}(0)=0.0144$ for $r=0.883$ and $g^{(2)}(0)=8.78\times 10^{-3}$ for $r=1.5$.
Figure~\ref{figure04}(b) indicates that $g^{(2)}(0)$ decreases as $r$ gets
larger
.
However, Fig.~\ref{figure04}(a) tells us that $P(1,1;r,\alpha)$ becomes maximum for $r=0.883$
and it decreases as $r$ gets larger than $r=0.883$.
Thus, Figs.~\ref{figure04}(a) and \ref{figure04}(b) mean that there is a trade-off.

\section{\label{section-imperfect-single-photon-detector}The efficiency and $g^{(2)}(0)$ for the case
where the single-photon detector is imperfect}
We estimate the effect of an imperfect single-photon detector on the efficiency and $g^{(2)}(0)$ of the heralded single-photon source.
We prepare a ``click/no click" detector with efficiency $\eta$
by the POVM (positive operator-valued measure) $\{\hat{M},\hat{\mbox{\boldmath $I$}}-\hat{M}\}$
where
\begin{equation}
\hat{M}
=
\eta
\sum_{k=1}^{\infty}
(1-\eta)^{k-1}|k\rangle\langle k|.
\end{equation}
The operators
$\hat{M}$ and $\hat{\mbox{\boldmath $I$}}-\hat{M}$ represent ``click'' and ``no click'' events, respectively,
where $\hat{\mbox{\boldmath $I$}}$ denotes the identity operator and $0\leq\eta\leq 1$
\cite{Resch2001,Kok2007,Akhlaghi2011}.
If $\eta=1$, we obtain $\hat{M}=|1\rangle\langle 1|$.
Here, we draw our attention to the fact $\mbox{tr}\hat{M}=1$.

The probability that we detect ``click" and $n$ heralded photons are emitted is given by
\begin{equation}
P_{\mbox{\scriptsize click}}(n;r,\alpha,\eta)
=
\eta
\sum_{k=1}^{\infty}
(1-\eta)^{k-1}
P(n,k;r,\alpha).
\label{P-click-0}
\end{equation}
Writing $n$ as the number of the emitted photons with ``click",
we can calculate the average of $n^{l}$ by
\begin{equation}
\langle n^{l}\rangle
=
\sum_{n=0}^{\infty}
n^{l}
P_{\mbox{\scriptsize click}}(n;r,\alpha,\eta).
\label{n-l-average-imperfect}
\end{equation}
Substituting Eq.~(\ref{n-l-average-imperfect}) into Eq.~(\ref{g2-0-definition}),
we can compute $g^{(2)}(0)$.

\begin{figure}
\begin{center}
\includegraphics[width=0.8\linewidth]{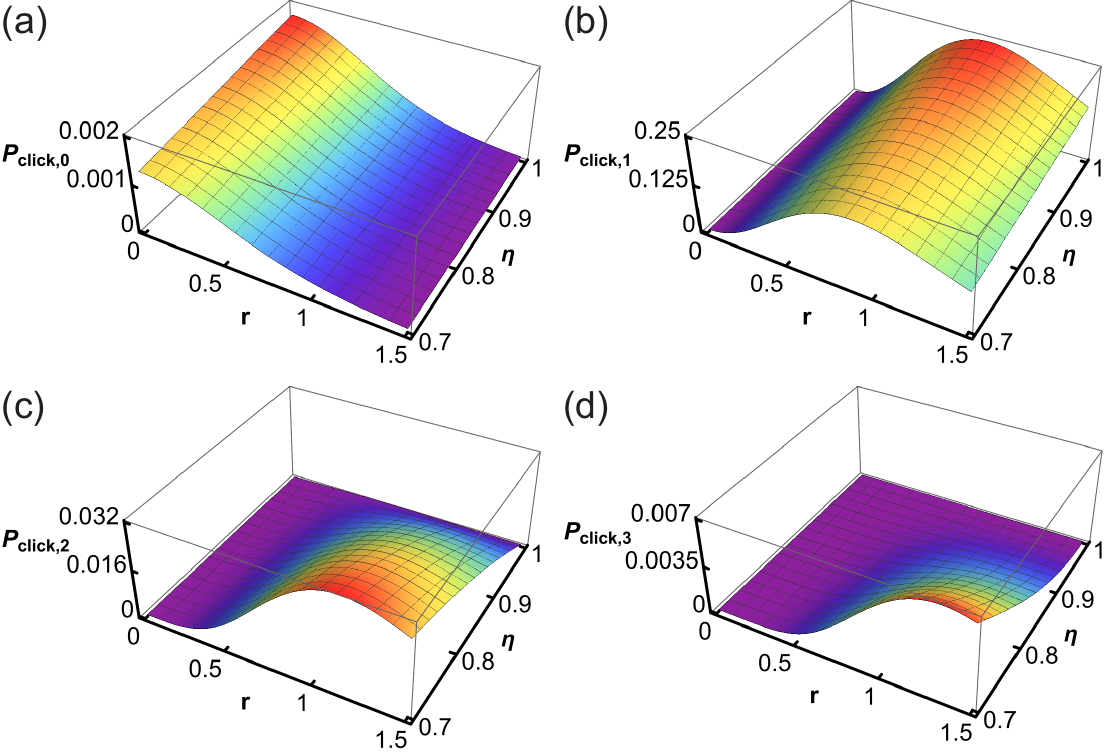}
\end{center}
\caption{
3D plots of $P_{\mbox{\scriptsize click}}(n;r,\alpha,\eta)$ given by Eq.~(\ref{P-click-0})
as functions of $r$ and $\eta$ with $\alpha=0.06$.
(a) $n=0$,
(b) $n=1$,
(c) $n=2$,
(d) $n=3$.
}
\label{figure05}
\end{figure}

\begin{figure}
\begin{center}
\includegraphics[width=0.65\linewidth]{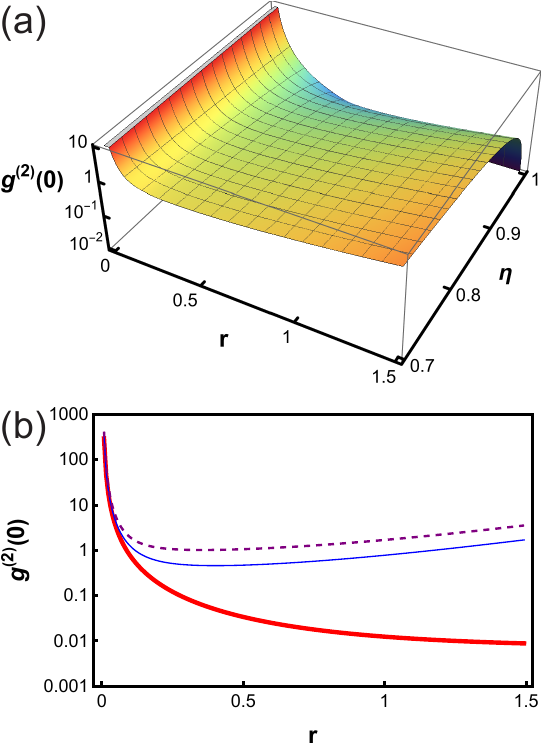}
\end{center}
\caption{(a)
A 3D plot of $g^{(2)}(0)$ given by
Eqs.~(\ref{g2-0-definition}) and (\ref{n-l-average-imperfect})
as a function of $r$ and $\eta$ with $\alpha=0.06$.
The vertical axis is displayed by the logarithmic scale.
(b)
Plots of $g^{(2)}(0)$ as functions of $r$ with $\alpha=0.06$.
The thick solid red, thin solid blue, and dashed purple curves represent $\eta=1.0$, $0.85$, and $0.7$, respectively.
The vertical axis is displayed by the logarithmic scale.}
\label{figure06}
\end{figure}

Figures~\ref{figure05}(a), (b), (c), and (d) show 3D plots of $P_{\mbox{\scriptsize click}}(n;r,\alpha,\eta)$
for $n=0, 1, 2$, and $3$
as functions of $r$ and $\eta$ with $\alpha=0.06$, respectively.
Looking at Fig.~\ref{figure05}(b), we note that the probability of obtaining the heralded single photon decreases
as $\eta$ gets smaller.
By contrast, looking at Figs.~\ref{figure05}(c) and (d), we recognize that the probabilities of emissions of two and three photons increase as $\eta$ gets smaller.
The probability $P_{\mbox{\scriptsize click}}(2;r,\alpha,\eta)$ reaches $0.03$ for $\eta=0.7$.
Thus, it is very unsafe to use this single-photon source
for the quantum key distribution under this circumstance
when the eavesdropper makes the photon number splitting attack.

Figures~\ref{figure06}(a) shows a 3D plot of
$g^{(2)}(0)$
as a function of $r$ and $\eta$ with $\alpha=0.06$.
Looking at Fig.~\ref{figure06}(a), we note that $g^{(2)}(0)$ increases drastically as $\eta$ gets smaller.
Figure~\ref{figure06}(b) shows plot of $g^{(2)}(0)$ as functions of $r$ with $\alpha=0.06$ and
$\eta=1.0$, $0.85$, and $0.7$.
Lookig at Fig.~\ref{figure06}(b), we note that $g^{(2)}(0)$ is larger than $0.1$ when $\eta$ is less than $0.85$.
Thus, we can conclude that the efficiency of the single-photon detector is crucial
for the quality of the heralded single-photon source.

Figure~\ref{figure06}(b) indicates that $g^{(2)}(0)$ decreases monotonically as $r$ gets larger for $\eta=1$.
Contrastingly, if $\eta=0.85$ and $0.7$, $g^{(2)}(0)$ increases as $r$ gets larger after rapid decreasing near $r=0$.
The reason for those observations is as follows.
To draw curves in Fig.~\ref{figure06}(b), we assume that the heralded photons are created from the two-mode squeezed state
given by Eq.~(\ref{two-mode-squeezed-state-with-alpha}) and $\alpha=0.06$.
Because $\alpha$ is small enough, we have an approximation with $\alpha=0$ for a while to examine the physical properties of the heralded photons.
Thus, the two-mode squeezed state is approximately given by
\begin{equation}
\hat{S}_{ab}(ir)|0\rangle_{a}|0\rangle_{b}
=
\mathrm{sech}\, r\sum_{n=0}^{\infty}
(-i \tanh r)^{n}|n\rangle_{a}|n\rangle_{b}.
\end{equation}
Hence, we can obtain a heralded single photon deterministically from the port $a$ if we detect a heralding single photon with $\eta=1$ from the port $b$.
For this case, we obtain $g^{(2)}(0)=0$ and we recognize that an ideal single-photon source is implemented.
Hence, the relation $g^{(2)}(0)\ll 1$ holds even if $r$ gets larger for $\eta=1$ for $\alpha=0.06$.
Here, we draw attention to the following facts.
When $r=0$ and $\eta=1$, the state of photons in the port $a$ is given by a coherent state $|\alpha\rangle$ and we obtain $g^{(2)}(0)=1$.
However, in the limit of $r\to 0$ and $r>0$,
we obtain $g^{(2)}(0)\gg 1$.
Thus, if we regard $g^{(2)}(0)$ as a function of $r$, it has a singularity at $r=0$ as mentioned in the caption of Fig.~\ref{figure03}(b).
Now, let us consider a case where $\eta<1$.
Then, we cannot perfectly distinguish between the states
$|0\rangle_{a}|0\rangle_{b}$, $|1\rangle_{a}|1\rangle_{b}$, $|2\rangle_{a}|2\rangle_{b}$, ...
and the performance of the port $a$ as the single-photon source declines.
Particularly, $\tanh r$ approaches unity
and the probability of observations of $|0\rangle_{a}|0\rangle_{b}$, $|1\rangle_{a}|1\rangle_{b}$, $|2\rangle_{a}|2\rangle_{b}$, ...
become equal to each other as $r$ gets larger,
and this makes $g^{(2)}(0)$ larger.
In fact, Refs.~\cite{Sempere-Llagostera2022,Azuma2024} reported that $g^{(2)}(0)$ increased as $r$ got larger where $\eta<1$.
In Ref.~\cite{Sempere-Llagostera2022}, Fig.~3(a) showed curves of $g^{(2)}(0)$ as functions of the heralding probability for $\eta=0.162$ and $\eta=0.296$.
In \cite{Azuma2024}, Fig.~9(a) showed a curve of $g^{(2)}(0)$ as a function of $r$ with $\eta=0.9$.

\section{\label{section-photonic-crystal-group-velocity-squeezing-parameter}
The group velocity of the light passing the photonic crystal and the squeezing parameter}
The derivation of the dispersion relation of the light passing the photonic crystal and
computation of its group velocity are important
when we generate the squeezed light with the photonic crystal composed of the nonlinear material.
In this section, we explain how to adjust the group velocity to achieve a desired squeezing parameter \cite{Azuma2022}.
Because of Eqs.~(\ref{definition-beta-parameter-0}) and (\ref{definition-squeezing-parameter-0}),
the squeezing parameter is inversely proportional to $v_{\mbox{\scriptsize g}}$.
We can obtain the dispersion relation $\omega=\omega(k)$ of the light passing the photonic crystal
shown in Fig.~\ref{figure01} using the following equations \cite{Azuma2008}:
\begin{eqnarray}
&&
\cos[(l_{\mbox{\scriptsize A}}+l_{\mbox{\scriptsize B}})k]
-
\cos(l_{\mbox{\scriptsize A}}K_{\mbox{\scriptsize A}})
\cos(l_{\mbox{\scriptsize B}}K_{\mbox{\scriptsize B}}) \nonumber \\
&&
\quad
+
\frac{K_{\mbox{\scriptsize A}}^{2}+K_{\mbox{\scriptsize B}}^{2}}
{2K_{\mbox{\scriptsize A}}K_{\mbox{\scriptsize B}}}
\sin(l_{\mbox{\scriptsize A}}K_{\mbox{\scriptsize A}})
\sin(l_{\mbox{\scriptsize B}}K_{\mbox{\scriptsize B}})
=
0,
\label{dispersion-relation-0}
\end{eqnarray}
\begin{equation}
K_{\mbox{\scriptsize A}}
=
\frac{\omega}{c}
\sqrt{\frac{\varepsilon_{\mbox{\scriptsize A}}}{\varepsilon_{0}}},
\quad
K_{\mbox{\scriptsize B}}
=
\frac{\omega}{c}
\sqrt{\frac{\varepsilon_{\mbox{\scriptsize B}}}{\varepsilon_{0}}},
\label{KA-KB-definition}
\end{equation}
where $\omega$ and $k$ represent the angular frequency and the wavenumber of the photon
in the photonic crystal, respectively.
We write the permittivities of vacuum, materials A and B as
$\varepsilon_{0}$,
$\varepsilon_{\mbox{\scriptsize A}}$, and $\varepsilon_{\mbox{\scriptsize B}}$, respectively.
The widths of the materials A and B are given by
$l_{\mbox{\scriptsize A}}$ and $l_{\mbox{\scriptsize B}}$, respectively.
We show derivation of Eqs.~(\ref{dispersion-relation-0}) and (\ref{KA-KB-definition})
in \ref{section-photonic-crystal-length-calibration}.

For a concrete example, we
assume that materials A and B are air and $\mbox{LiNbO}_{3}$, respectively.
The relative permittivity of air is given by
$\varepsilon_{\mbox{\scriptsize A}}/\varepsilon_{0}=1$.
The refractive index of $\mbox{LiNbO}_{3}$
is given by
$n_{\mbox{\scriptsize B}}=2.2$
and
we obtain the relative permittivity
$\varepsilon_{\mbox{\scriptsize B}}/\varepsilon_{0}=n_{\mbox{\scriptsize B}}^{2}$
\cite{Smith1976}.
As reported in Ref.~\cite{Li2020,Jiang2020,Larocque2024},
we place
$l_{\mbox{\scriptsize A}}=l_{\mbox{\scriptsize B}}=5.0\times 10^{-7}$ m.
The $\mbox{LiNbO}_{3}$ is transparent for the light beam where wavelength is between
$\lambda=4.0\times 10^{-7}$ m
and
$\lambda=4.8\times 10^{-6}$ m
\cite{Jundt1997,Gayer2008}.

\begin{figure}
\begin{center}
\includegraphics[width=0.65\linewidth]{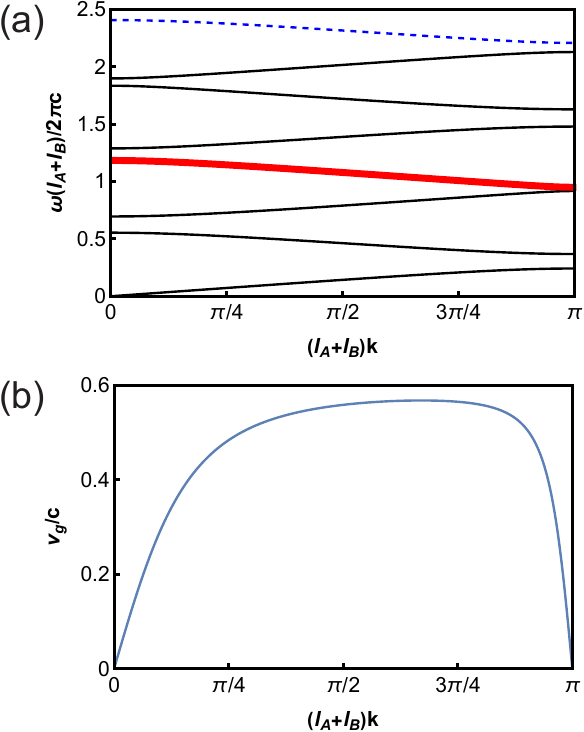}
\end{center}
\caption{(a)
The dispersion relation for the photonic crystal,
that is,
the angular frequency as a function of the wavenumber.
The thick red and dashed blue curves represent the fourth and eighth conduction bands from the bottom.
(b)
The group velocity of the fourth conduction band from the bottom.
At both ends of the graph, $v_{\mbox{\scriptsize g}}/c$ is equal to zero.}
\label{figure07}
\end{figure}

Figure~\ref{figure07}(a) shows a plot of the dispersion relation of the light in the photonic crystal.
If the wavenumber is equal to $k_{0}$,
the group velocity $v_{\mbox{\scriptsize g}}(k_{0})$ is given by:
\begin{equation}
v_{\mbox{\scriptsize g}}(k_{0})
=
\left|
\frac{d\omega}{dk}
\right|_{k=k_{0}}.
\end{equation}
Figure~\ref{figure07}(b) shows a plot of the group velocity $v_{\mbox{\scriptsize g}}/c$
of the fourth conduction band from the bottom as a function of the wavenumber
$(l_{\mbox{\scriptsize A}}+l_{\mbox{\scriptsize B}})k$.
In Fig.~\ref{figure07}(b), the left end of the graph means $k=0$ and $v_{\mbox{\scriptsize g}}=0$.
At this point, the angular frequency and the wavelength of the signal light are given by:
$\omega_{\mbox{\scriptsize s}}(l_{\mbox{\scriptsize A}}+l_{\mbox{\scriptsize B}})/
(2\pi c)=1.18$
and
$\lambda_{\mbox{\scriptsize s}}=8.45\times 10^{-7}$ m.
Because the angular frequency of the pump photons is given by
$\omega_{\mbox{\scriptsize p}}=2\omega_{\mbox{\scriptsize s}}$,
we obtain
$\lambda_{\mbox{\scriptsize p}}=4.22\times 10^{-7}$ m
and
$\omega_{\mbox{\scriptsize p}}(l_{\mbox{\scriptsize A}}+l_{\mbox{\scriptsize B}})/
(2\pi c)=2.37$.
Thus, the pump light is included in the eighth conduction band from the bottom.
In those concrete examples,
the signal and pump light beams are within the range of the wavelength
of transparent light for $\mbox{LiNbO}_{3}$.

The second-order nonlinear optical susceptibility of $\mbox{LiNbO}_{3}$
is given by
$\chi^{(2)}=\varepsilon_{0}\tilde{\chi}^{(2)}$
and
$\tilde{\chi}^{(2)}=25.2\times 10^{-12}$ $\mbox{mV}^{-1}$
\cite{Shoji1997,Kawase2002,Schiek2012}.
We assume that the total length of the photonic crystal is given by
$l=5.0\times 10^{-5}$ m.
We can write the squeezing parameter $\zeta$ as $\zeta=\beta l$, where
\begin{equation}
\beta
=
\frac{\omega_{\mbox{\scriptsize s}}A\tilde{\chi}^{(2)}}{v_{\mbox{\scriptsize g}}}.
\label{beta-definition-dash}
\end{equation}
Then, the amplitude of the pump light is given by
\begin{eqnarray}
A
&=&
\frac{v_{\mbox{\scriptsize g}}\zeta}
{\omega_{\mbox{\scriptsize s}}\tilde{\chi}^{(2)}l}
\nonumber \\
&=&
1.07\times 10^{8}
\frac{v_{\mbox{\scriptsize g}}\zeta}{c}
\quad
\mbox{Vm}^{-1}.
\end{eqnarray}
Here, we estimate $A$ practically.
The relation between the intensity
$I$ ($\mbox{Wm}^{-2}$)
and
amplitude $A$ ($\mbox{Vm}^{-1}$)
of the electric field of the pump laser is given by
\begin{equation}
I
=
\frac{1}{2}
\varepsilon_{0}cnA^{2},
\label{intensity-relation-A}
\end{equation}
where $n$ denotes the refractive index of the vacuum.
The standard specification of a semiconductor laser for commercial use is given by the radiant flux
$W=0.03$ W
and the radius of the laser beam
$d=5.0\times 10^{-6}$ m.
From Eq.~(\ref{intensity-relation-A}),
\begin{equation}
I
=
\frac{W}{\pi d^{2}}
\end{equation}
and the approximation $n=1$,
we get
$A=5.36\times 10^{5}$ $\mbox{Vm}^{-1}$.
Thus, we have
\begin{equation}
\frac{v_{\mbox{\scriptsize g}}\zeta}{c}=5.01\times 10^{-3}.
\end{equation}
If we want to attain $\zeta=1$, we must obtain
$v_{\mbox{\scriptsize g}}/c=5.01\times 10^{-3}$.
Looking at Fig.~\ref{figure07}(b),
we note that a range
$0\leq v_{\mbox{\scriptsize g}}/c\leq 5.01\times 10^{-3}$
corresponds to a range
$0\leq (l_{\mbox{\scriptsize A}}+l_{\mbox{\scriptsize B}})k\leq 3.15\times 10^{-3}$.
We do not consider a range around $(l_{\mbox{\scriptsize A}}+l_{\mbox{\scriptsize B}})k=\pi$
because the curve of $v_{\mbox{\scriptsize g}}/c$ near this range is steeper than that near $(l_{\mbox{\scriptsize A}}+l_{\mbox{\scriptsize B}})k=0$
and it is less suitable for experiments.
Then, looking at Fig.~\ref{figure07}(a),
we note that the above range corresponds to
$\Delta\omega=7.86\times 10^{9}$ $\mbox{s}^{-1}$.
This implies that we need to adjust the frequency $\nu$ with a precision
$\Delta\nu=\Delta\omega/(2\pi)=1.25\times 10^{9}$ Hz.
Hence, we must set the frequency of the signal light with a ratio $3.52\times 10^{-6}$
because of
$\nu_{\mbox{\scriptsize s}}
=
\omega_{\mbox{\scriptsize s}}/(2\pi)=3.55\times 10^{14}$ Hz.

Here, we consider the following problem.
So far, we compute the squeezing parameter by the relation $\zeta=\beta l$
where $\beta$ is given by Eq.~(\ref{beta-definition-dash}).
Thus, the squeezing parameter is in proportion to the time duration
$\tau=l/v_{\mbox{\scriptsize g}}$.
Thus, $\tau$ is the amount of time the photons spend on passing through the photonic crystal.
However, strictly speaking, $\tau$ must be equal to the time duration that the photons need
for traveling through the layers of $\mbox{LiNbO}_{3}$ only.

Now, we introduce a ratio $P_{\mbox{\scriptsize A}}:P_{\mbox{\scriptsize B}}$,
where $P_{\mbox{\scriptsize A}}$ and $P_{\mbox{\scriptsize B}}$ are the probabilities
that the photons exist in the layers of the materials A and B, respectively.
Let $\tau_{\mbox{\scriptsize A}}$ and $\tau_{\mbox{\scriptsize B}}$
be time durations the photons take to travel through the materials A and B, respectively.
Then, $\tau=\tau_{\mbox{\scriptsize A}}+\tau_{\mbox{\scriptsize B}}$
and
$\tau_{\mbox{\scriptsize A}}:\tau_{\mbox{\scriptsize B}}
=
P_{\mbox{\scriptsize A}}:P_{\mbox{\scriptsize B}}$
hold.
Hence, we must replace $\tau=l/v_{\mbox{\scriptsize g}}$ with
\begin{equation}
\tau'
=
\frac{l}{v_{\mbox{\scriptsize g}}}
\frac{P_{\mbox{\scriptsize B}}}{P_{\mbox{\scriptsize A}}+P_{\mbox{\scriptsize B}}},
\end{equation}
where $\tau'$ represents the total time duration the photons take to travel through the layers of the material B.
In other words, we must calibrate the total length of the photonic crystal by changing from $l$
to $l(P_{\mbox{\scriptsize A}}+P_{\mbox{\scriptsize B}})/P_{\mbox{\scriptsize B}}$.
In \ref{section-photonic-crystal-length-calibration},
we numerically calculate $P_{\mbox{\scriptsize A}}:P_{\mbox{\scriptsize B}}$
for a concrete example,
$\lambda_{\mbox{\scriptsize s}}=8.45\times 10^{-7}$ m,
and we obtain
$P_{\mbox{\scriptsize A}}:P_{\mbox{\scriptsize B}}=1:1.32$.
Thus, we must change the total length of the photonic crystal
from $l$ to $1.76\times l$,
that is,
from $l=5.0\times 10^{-5}$ m to $l=8.79\times 10^{-5}$ m
to realize the squeezing parameter $\zeta=1$.

In this section, we adjust physical parameters to attain $\zeta=r=1$.
It is an example and our discussion is not limited to this.
The squeezing parameter $r=1$ corresponds to $8.69$ dB and it is a challenging value.
In Refs.~\cite{Glorieux2010,Kim2018,Sim2025}, $9.2$ dB, $5.4$ dB, and $5.0$ dB for intensity-difference squeezing from four-wave mixing were reached, respectively.

\section{\label{section-conclusions-discussions}Conclusions and discussions}
In this paper, we proposed how to implement the heralded single-photon source with one-dimensional photonic crystals.
In this proposal, first, we transformed incident coherent light into squeezed light with the photonic crystal
that was composed of the material with the large second-order nonlinear optical susceptibility $\chi^{(2)}$.
Second, by injecting two squeezed light beams into the 50-50 beam splitter,
we obtained the two-mode squeezed state.
Third, utilizing the entanglement between two modes of the photons,
we constructed the heralded single-photon source.

The efficiency of the single photons of the source is more than $0.2$ per pump photon.
Because of this point, our proposal is superior to conventional methods realized by the SPDC.
The disadvantage of our method is that
we have to adjust the frequencies of the pump and signal light beams precisely to obtain a large squeezing parameter.
According to our calculations for the concrete example,
we must prepare the signal light beam of  frequency
$\nu_{\mbox{\scriptsize s}}=3.55\times 10^{14}$ Hz
with an accuracy of $\Delta\nu_{\mbox{\scriptsize s}}=1.25\times 10^{9}$ Hz.
This setting requires the precision of the frequency with the ppm (parts per million),
and it is achievable experimentally.
Therefore, we can conclude that our proposed method is practical.

In this paper, we investigated a method to strengthen a quantum characteristic by photonic crystals.
Although the slow group velocity of the light is caused by classical electromagnetism,
it affects the quantum nature of states of photons.
We think that the photonic crystals have wide applications according to this policy.

\appendix

\section{\label{section-photonic-crystal-length-calibration}
The ratio of probabilities that the photons exist in the layers of the materials A and B}
We compute a ratio
$P_{\mbox{\scriptsize A}}:P_{\mbox{\scriptsize B}}$,
where $P_{\mbox{\scriptsize A}}$ and $P_{\mbox{\scriptsize B}}$
are
the probabilities that the photons are localized in the layers of the materials A and B,
respectively
\cite{Sakoda2005,Azuma2008}.
The wave equation of the electromagnetic field in a one-dimensional photonic crystal is given by
\begin{equation}
\frac{\partial^{2}}{\partial x^{2}}
E(x,t)
-
\mu_{0}\varepsilon(x)
\frac{\partial^{2}}{\partial t^{2}}
E(x,t)
=
0,
\label{wave-equation-electric-field-0}
\end{equation}
\begin{equation}
\frac{\partial}{\partial x}
E(x,t)
=
-
\frac{\partial}{\partial t}
B(x,t)
\label{electric-magnetic-field}
\end{equation}
\begin{equation}
\varepsilon(x+l_{\mbox{\scriptsize A}}+l_{\mbox{\scriptsize B}})
=
\varepsilon(x)
\quad
\mbox{$-\infty<x<\infty$},
\label{varepsilon-periodic}
\end{equation}
\begin{equation}
\varepsilon(x)
=
\left\{
\begin{array}{ll}
\varepsilon_{\mbox{\scriptsize A}}(>0) & 0\leq x< l_{\mbox{\scriptsize A}},\\
\varepsilon_{\mbox{\scriptsize B}}(>0) & l_{\mbox{\scriptsize A}}\leq x< l_{\mbox{\scriptsize A}}+l_{\mbox{\scriptsize B}}.\\
\end{array}
\right.
\label{varepsilon-medium-A-B}
\end{equation}
Moreover, we assume that the injected pulse into the photonic crystal is monochromatic as
\begin{equation}
E(x,t)=E(x)e^{-i\omega t}.
\label{wave-equation-electric-field-1}
\end{equation}
Accordingly, we can rewrite the wave equation as
\begin{equation}
\left[
\frac{\partial^{2}}{\partial x^{2}}
+
\frac{\varepsilon(x)}{\varepsilon_{0}}
\left(
\frac{\omega}{c}
\right)^{2}
\right]
E(x)
=
0,
\label{wave-equation-electric-field-2}
\end{equation}
where we use
$c=1/\sqrt{\varepsilon_{0}\mu_{0}}$.

Let us solve
Eqs.~(\ref{varepsilon-periodic}), (\ref{varepsilon-medium-A-B}),
and (\ref{wave-equation-electric-field-2}).
Solutions of the region A ($0<x<l_{\mbox{\scriptsize A}}$)
and the region B ($l_{\mbox{\scriptsize A}}<x<l_{\mbox{\scriptsize A}}+l_{\mbox{\scriptsize B}}$)
for
Eqs.~(\ref{varepsilon-medium-A-B})
and (\ref{wave-equation-electric-field-2})
are given by
\begin{equation}
E_{j}(x)
=
C_{j+}\exp(iK_{j}x)
+
C_{j-}\exp(-iK_{j}x)
\quad
\mbox{for $j=\mbox{A, B}$},
\end{equation}
where $K_{\mbox{\scriptsize A}}$ and $K_{\mbox{\scriptsize B}}$ are given by
Eq.~(\ref{KA-KB-definition}).
Because $\varepsilon(x)$ is periodic,
we can apply Bloch's theorem to the continuity condition of the solution.
\\
\noindent
\textbf{Bloch's theorem}
\\
We write the one-dimensional electric field whose wavenumber is equal to $k$ as $E(x)$.
Suppose $E(x)$ is a solution of Eqs.~(\ref{varepsilon-periodic})
and (\ref{wave-equation-electric-field-2}).
Then, $E(x)$ satisfies the following relations:
\begin{equation}
E(x)=e^{ikx}u_{k}(x),
\end{equation}
\begin{equation}
u_{k}(x+l_{\mbox{\scriptsize A}}+l_{\mbox{\scriptsize B}})
=
u_{k}(x)
\quad
\mbox{for $-\infty<x<\infty$},
\end{equation}
\begin{equation}
-
\frac{\pi}{l_{\mbox{\scriptsize A}}+l_{\mbox{\scriptsize B}}}
\leq
k
\leq
\frac{\pi}{l_{\mbox{\scriptsize A}}+l_{\mbox{\scriptsize B}}}.
\end{equation}

The continuity conditions of
$E_{\mbox{\scriptsize A}}(x)$
and
$E_{\mbox{\scriptsize B}}(x)$ are given by
\begin{eqnarray}
E_{\mbox{\scriptsize A}}(0)
&=&
E_{\mbox{\scriptsize B}}(0), \nonumber \\
\left.
\frac{d}{dx}
E_{\mbox{\scriptsize A}}(x)
\right|_{x=0}
&=&
\left.
\frac{d}{dx}
E_{\mbox{\scriptsize B}}(x)
\right|_{x=0}, \nonumber \\
E_{\mbox{\scriptsize A}}(l_{\mbox{\scriptsize A}})
&=&
e^{ik(l_{\mbox{\tiny A}}+l_{\mbox{\tiny B}})}
E_{\mbox{\scriptsize B}}(-l_{\mbox{\scriptsize B}}), \nonumber \\
\left.
\frac{d}{dx}
E_{\mbox{\scriptsize A}}(x)
\right|_{x=l_{\mbox{\tiny A}}}
&=&
e^{ik(l_{\mbox{\tiny A}}+l_{\mbox{\tiny B}})}
\left.
\frac{d}{dx}
E_{\mbox{\scriptsize B}}(x)
\right|_{x=-l_{\mbox{\tiny B}}}.
\end{eqnarray}
We can rewrite the above equations as
\begin{equation}
M
\left(
\begin{array}{c}
C_{\mbox{\tiny A}+} \\
C_{\mbox{\tiny A}-} \\
C_{\mbox{\tiny B}+} \\
C_{\mbox{\tiny B}-}
\end{array}
\right)
=
0,
\end{equation}
\begin{equation}
M
=
\left(
\begin{array}{cccc}
1 & 1 & -1 & -1 \\
K_{\mbox{\scriptsize A}} & -K_{\mbox{\scriptsize A}} & -K_{\mbox{\scriptsize B}} & K_{\mbox{\scriptsize B}} \\
P & 1/P & -Q/R & -QR \\
K_{\mbox{\scriptsize A}}P & -K_{\mbox{\scriptsize A}}/P & -K_{\mbox{\scriptsize B}}Q/R & K_{\mbox{\scriptsize B}}QR
\end{array}
\right),
\end{equation}
\begin{eqnarray}
P
&=&
\exp(iK_{\mbox{\scriptsize A}}l_{\mbox{\scriptsize A}}), \nonumber \\
Q
&=&
\exp[ik(l_{\mbox{\scriptsize A}}+l_{\mbox{\scriptsize B}})], \nonumber \\
R
&=&
\exp(iK_{\mbox{\scriptsize B}}l_{\mbox{\scriptsize B}}).
\end{eqnarray}
We obtain Eq.~(\ref{dispersion-relation-0}) which gives us the dispersion relation
because of $\mbox{det}M=0$.

Because the number of photons in a specific volume is in proportion to the energy contained by it,
we can write the ratio as
\begin{eqnarray}
&&
P_{\mbox{\scriptsize A}}:P_{\mbox{\scriptsize B}} \nonumber \\
&=&
\frac{\omega}{2\pi}
\int_{0}^{2\pi/\omega}dt
S
\int_{0}^{l_{\mbox{\tiny A}}}dx
\frac{1}{2}
\left[
\varepsilon_{\mbox{\scriptsize A}}
E_{\mbox{\scriptsize A}}^{2}(x,t)
+
\frac{1}{\mu_{0}}B_{\mbox{\scriptsize A}}^{2}(x,t)
\right] \nonumber \\
&&
:
\frac{\omega}{2\pi}
\int_{0}^{2\pi/\omega}dt
S
\int_{l_{\mbox{\tiny A}}}^{l_{\mbox{\tiny A}}+l_{\mbox{\tiny B}}}dx
\frac{1}{2}
\left[
\varepsilon_{\mbox{\scriptsize B}}
E_{\mbox{\scriptsize B}}^{2}(x,t)
+
\frac{1}{\mu_{0}}B_{\mbox{\scriptsize B}}^{2}(x,t)
\right] \nonumber \\
&=&
\int_{0}^{2\pi/\omega}dt
\int_{0}^{l_{\mbox{\tiny A}}}dx
\left[
\frac{\varepsilon_{\mbox{\scriptsize A}}}{\varepsilon_{0}c^{2}}
E_{\mbox{\scriptsize A}}^{2}(x,t)
+
B_{\mbox{\scriptsize A}}^{2}(x,t)
\right] \nonumber \\
&&
:
\int_{0}^{2\pi/\omega}dt
\int_{l_{\mbox{\tiny A}}}^{l_{\mbox{\tiny A}}+l_{\mbox{\tiny B}}}dx
\left[
\frac{\varepsilon_{\mbox{\scriptsize B}}}{\varepsilon_{0}c^{2}}
E_{\mbox{\scriptsize B}}^{2}(x,t)
+
B_{\mbox{\scriptsize B}}^{2}(x,t)
\right],
\label{material-AB-probability-ratio}
\end{eqnarray}
where $S$ denotes the cross-sectional area of the photonic crystal
and we take an average of the energy over time $t$.
The electromagnetic fields
$E_{j}(x,t)$ and $B_{j}(x,t)$ for $j=\mbox{A, B}$ in Eq.~(\ref{material-AB-probability-ratio}) are given by
\begin{eqnarray}
E_{j}(x,t)
&=&
\mbox{Re}
\{
[C_{j+}\exp(iK_{j}x)
+
C_{j-}\exp(-iK_{j}x)]\exp(-i\omega t)
\}, \nonumber \\
B_{j}(x,t)
&=&
\mbox{Re}
\left\{
\frac{K_{j}}{\omega}
[C_{j+}\exp(iK_{j}x)
-
C_{j-}\exp(-iK_{j}x)]\exp(-i\omega t)
\right\} \nonumber \\
&&
\quad
\mbox{for $j=\mbox{A, B}$}.
\end{eqnarray}

As explained in Sec.~\ref{section-photonic-crystal-group-velocity-squeezing-parameter},
at the left end of the fourth conduction band,
we obtain $v_{\mbox{\scriptsize g}}=0$,
$\omega_{\mbox{\scriptsize s}}(l_{\mbox{\scriptsize A}}+l_{\mbox{\scriptsize B}})/
(2\pi c)=1.18$,
and $k=0$.
Substitution of this
$\omega_{\mbox{\scriptsize s}}$
for the signal photon
and specific values of the physical parameters
$\varepsilon_{\mbox{\scriptsize A}}$,
$\varepsilon_{\mbox{\scriptsize B}}$,
$l_{\mbox{\scriptsize A}}$,
and $l_{\mbox{\scriptsize B}}$
into
Eq.~(\ref{material-AB-probability-ratio}) provides us with
$P_{\mbox{\scriptsize A}}:P_{\mbox{\scriptsize B}}=1:1.32$.

\section*{Acknowledgment}
This work was supported by MEXT Quantum Leap Flagship Program Grant No. JPMXS0120351339.

\end{document}